\documentclass[
	aps, 
	prd,  
	notitlepage, 
    floats, 
    floatfix, 
    onecolumn,
	amsmath, 
	amssymb, 
	amsfonts, 
	eqsecnum,
	superscriptaddress,
	showpacs, 
	showkeys,
	nofootinbib,
 	longbibliography,
]{revtex4-2}

\usepackage{graphicx}
\usepackage{dcolumn}
\usepackage{bm}

\usepackage{color}

\def\comment#1{}

\begin{document}


\title{Exploring the nature of black hole and gravity with an imminent merging binary of supermassive black holes}

\author{Zhong Xingyu}
\affiliation{Shanghai Astronomical Observatory, Chinese Academy of Sciences, Shanghai 200030, China}
\affiliation{School of Astronomy and Space Science, University of Chinese Academy of Sciences, Beijing 100049, China}

\author{Han Wenbiao}
\email{corresponding author: wbhan@shao.ac.cn}
\affiliation{Shanghai Astronomical Observatory, Chinese Academy of Sciences, Shanghai 200030, China}
\affiliation{Hangzhou Institute for Advanced Study, University of Chinese Academy of Sciences, Hangzhou 310124, China}
\affiliation{Taiji Laboratory for Gravitational Wave Universe (Beijing/Hangzhou), University of Chinese Academy of Sciences, Beijing 100049, China}
\affiliation{School of Astronomy and Space Science, University of Chinese Academy of Sciences, Beijing 100049, China}
\affiliation{Shanghai Frontiers Science Center for  Gravitational Wave Detection, 800 Dongchuan Road, Shanghai 200240, China}

\author{Luo Ziren} 
\affiliation{Taiji Laboratory for Gravitational Wave Universe (Beijing/Hangzhou), University of Chinese Academy of Sciences, Beijing 100049, China}
\affiliation{Hangzhou Institute for Advanced Study, University of Chinese Academy of Sciences, Hangzhou 310124, China}
\affiliation{Institute of Mechanics, Chinese Academy of Sciences, Beijing 100190, China} 
\affiliation{Key Laboratory of Gravitational Wave Precision Measurement of Zhejiang Province, Hangzhou 310124, China}

\author{Wu Yueliang} 
\affiliation{Taiji Laboratory for Gravitational Wave Universe (Beijing/Hangzhou), University of Chinese Academy of Sciences, Beijing 100049, China}
\affiliation{Hangzhou Institute for Advanced Study, University of Chinese Academy of Sciences, Hangzhou 310124, China}
\affiliation{Institute of Theoretical Physics, Chinese Academy of Sciences, Beijing 100190} 
\affiliation{International Centre for Theoretical Physics Asia-Pacific, UCAS, Beijing 100190}

\date{\today}

\begin{abstract}
A supermassive binary black-hole candidate SDSS J1430+2303 reported recently motivates us to investigate an imminent binary of supermassive black holes as potential gravitational wave source, the radiated gravitational waves at the end of the merger are shown to be in the band of space-borne detectors. We provide a general analysis on the required detecting sensitivity needed for probing such type gravitational wave sources and make a full discussion by considering two typically designed configurations of space-borne antennas. If a source is so close, it is possible to be detected with Taiji pathfinder-plus which is proposed to be an
extension for the planned Taiji pathfinder by just adding an additional satellite to the initial two satellites. The gravitational wave detection on such kind of source enables us to explore the properties of supermassive black holes and the nature of gravity.
\begin{description}
\item[PACS numbers]
04.70.Bw, 04.80.Nn, 95.10.Fh

\end{description}
\end{abstract}

\maketitle
\section{Introduction}
So far, a total of 90 gravitational-wave (GW) events have been announced by the LIGO-Virgo-KAGRA (LVK) collaboration after the third observing run (O3) \cite{lvkGWTC3}. GW astronomy has begun to embrace the explosion of scientific outputs. The ground-based detectors focus on GW signals in the high-frequency band, which originate mainly from compact stellar objects. Future space-borne detectors such as LISA\cite{elisa}, Taiji\cite{hu2017taiji} and Tianqin\cite{2016CQGra..33c5010L}, will observe GWs in the low-frequency band from 0.1 mHz to 1 Hz. All three detectors are expected to be launched in the 2030s, opening the window for future observations of GW sources involving supermassive black holes (SMBHs).

LISA consists of three satellites orbiting the sun and with an average inter-satellite separation of $2.5\times10^6$ km. In the case of Taiji, the orbital configuration is similar but the interferometer arm length is $3\times10^6$ km. Moreover, the sensing noise is about 8 pm/$\sqrt{\rm Hz}$ and the test-mass force noise is 6 fN/$\sqrt{\rm Hz}$ \cite{hu2017taiji}. Due to the periodic motions in orbit around the Sun, space-based detectors will be able to observe long-duration GW signals at different positions and orientations.

Supermassive black hole binaries (SMBHBs) are products of galaxy mergers in the hierarchical universe \cite{begelman1980massive}. Early studies of SMBHBs always assumed that they had circular orbits for convenience. It is expected that environmental conditions such as gas friction and star scattering may reduce the eccentricity of the orbit. So, it will make the expected merger beyond Hubble time. Considering a highly eccentric trajectory \cite{Armitage_2005, Berczik_2006,2011ApJ...729...13C}, the secondary black hole intersects the orbital periastron point, where significant energy and angular momentum are dissipated by GW emission, accelerating the coalescence. In addition, at the final stage of their orbital evolution, the GW radiation drives the motion of the binary and produces the strongest siren \cite{1976ApJ...204L...1T,1994MNRAS.269..199H,2003ApJ...583..616J} which can be detected by GW observatories. 

In the case of SMBHBs as GW sources, the huge mass will lead to a GW signal in the low frequency range, around mHz, therefore, SMBHSs are very important sources for space-borne detectors. The detection of an SMBHB signal will be a unique way to test general relativity (GR) and the nature of black holes. Recently, it has been reported that the Seyfert 1-type galaxy SDSS J1430+2303 hosts an SMBHB candidate \cite{2022arXiv220111633J} located at a redshift 0.08105 \cite{2011ApJS..195...13O} (hereafter, we will refer to this SMBHB as J1430+2303). The optical luminosity variations of the galaxy SDSS J1430+2303 were monitored using the $g$ and $r$-band light curves of the Zwicky Transient Facility (ZTF) \cite{Bellm_2018}, which exhibit an oscillation pattern since early 2019. Ref. \cite{2022arXiv220111633J} predicts that the merger time of J1430+2303 is about three years by fitting the trajectory models with the optical light curves. This could therefore be the first observation of an SMBHB coalescence event in human history. Ref. \cite{VLBI_J1430} uses the very long baseline interferometry (VLBI) imaging method to observe (in late February and early March 2022) J1430+2303 in milliarcsecond-level high resolution.
The imaging highlights the existence of a very compact component with a temperature above $10^8K$ and an unresolved morphology with a size smaller than $0.8$ pc in the core of the AGN.  Furthermore, it indicates that there are no radio bursts, which denotes that J1430+2303 has not yet merged and it does not mention the merger time. Consequently, it is not yet confirmed that J1430+2303 could be an immanent binary, but it encourages us to consider potential immanent binaries of supermassive black holes.

Although a GW source such as J1430+2303 is an ideal source for space-based detectors, if the merger time is expected to be around 2025, Taiji and LISA will not be able to detect this event.
The expected signal strength for SMBHBs such as J1430+2303 suggests that it could still be detected by a low-level interferometer with a sensing noise of a few hundred pm/$\sqrt{\rm Hz}$ and acceleration noise of several hundred fN /$\sqrt{\rm Hz}$. The Taiji project has planned a pathfinder launch around 2025 with a pair of satellites using a million km long arm, with a lower requirement proposed for a sensing noise of 200 pm/$\sqrt{\rm Hz}$ and a test-mass force noise of 600 fN/$\sqrt{\rm Hz}$ \cite{LUO2020102918}. 
In the meantime, it has been suggested that an additional satellite be added to the initial configuration, called Taiji pathfinder plus. We will see that this Taiji pathfinder plus (hereafter called pathfinder plus-A) can detect GWs from SMBHBs such as J1430+2303 with a signal-to-noise ratio (SNR) of about 10. In addition, the optimum pathfinder configuration should achieve noise levels of 100 pm/$\sqrt{\rm Hz}$ and 60 fN/$\sqrt{\rm Hz}$. If this objective is reached, the proposed Taiji pathfinder plus would attain a much higher sensitivity. We will refer to this pathfinder plus configuration as the Taiji pathfinder plus-B. In that case, it will be possible to detect a GW signal from an SMBHB, such as J1430+2303, with an SNR sufficiently large to extract the necessary information about SMBHBs and to accomplish a series of related scientific tasks.

In this paper we will discuss the potential to detect the J1430+2303 type SMBHB through Taiji plus-A and Taiji plus-B. It should be noted that although J1430+2303 has not yet been confirmed as an immanent binary, all the calculations presented in this paper based on the J1430+2303 hold for all potential imminent merging binary SMBHs, which will make sense for the scientific goals of the Taiji project. We will show that the expected SNR is large enough to be detected by pathfinder plus-A, and thus by pathfinder plus-B, since the latter has a better sensitivity.
In addition, we will explore in detail the ability to address some key scientific issues related to the J1430+2303 type SMBHBs with pathfinder plus-B.  This paper is organized as follows, in section 2 we will discuss the GW signal from J1430+2303 and measure the key parameters considering a detection for Taiji plus-A, Taiji plus-B, and the final Taiji pathfinder configuration. In section 3, we will focus on some scientific objectives that could be investigated using the detection of a GW signal from J1430+2303. The last section will summarize this work and discuss the detecting and measuring of the ability of three kinds of Taiji detectors on this type of source.

\section{An imminent merging binary as a GW source}
SDSS J1430+2303 consists of a primary SMBH with a mass of $1.6\times10^8M_{\odot}$ and a secondary SMBH with a mass of $4\times10^7M_{\odot}$, with an inclined and highly eccentric trajectory. Despite this, we can use a waveform derived for a circular orbit to approximate the expected signal, since the contribution of the high eccentricity is mainly distributed in the low frequency band which is outside the detection range of the three types of detectors considered here. Here we use ``SEOBNRE" \cite{cao2017waveform} model to compute the waveform with eccentricity in the time domain and the ``IMRPhenomD" method \cite{IMRD} which is a frequency-domain phenomenological model for the GW signal from the inspiral, merger and ringdown (IMR) of black-hole binaries to get the frequency domain signal approximate.

\begin{figure}
\centering
\includegraphics[width=0.49\textwidth]{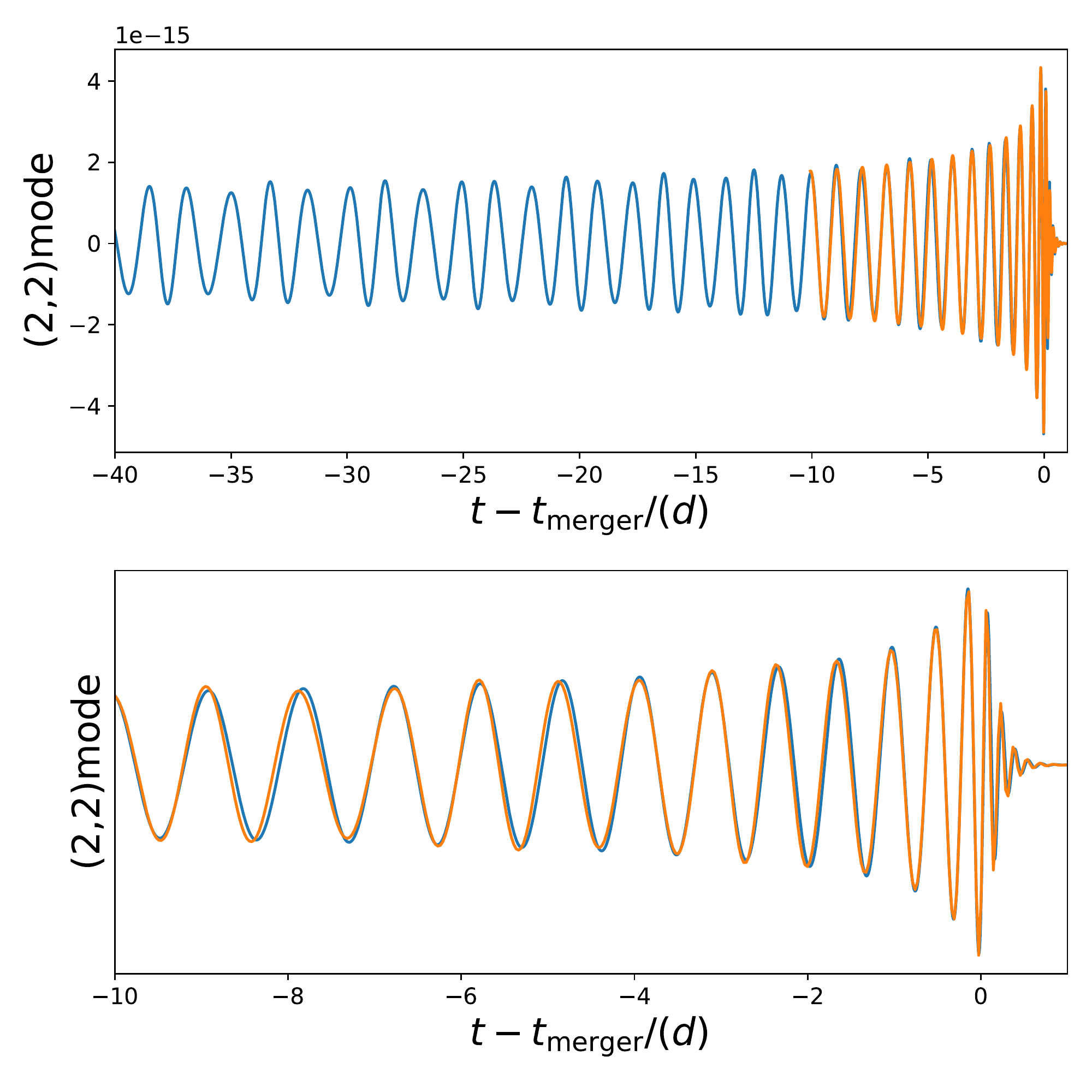}
\caption{\label{td_seobnre} Time domain GW waveform for a null eccentricity $e=0$ in orange, and for $e=0.9$ in blue . The $(2,2)$ mode refers to the strain mode with $l=m=2$. }
\end{figure}

\begin{figure}
\centering
\includegraphics[width=0.49\textwidth]{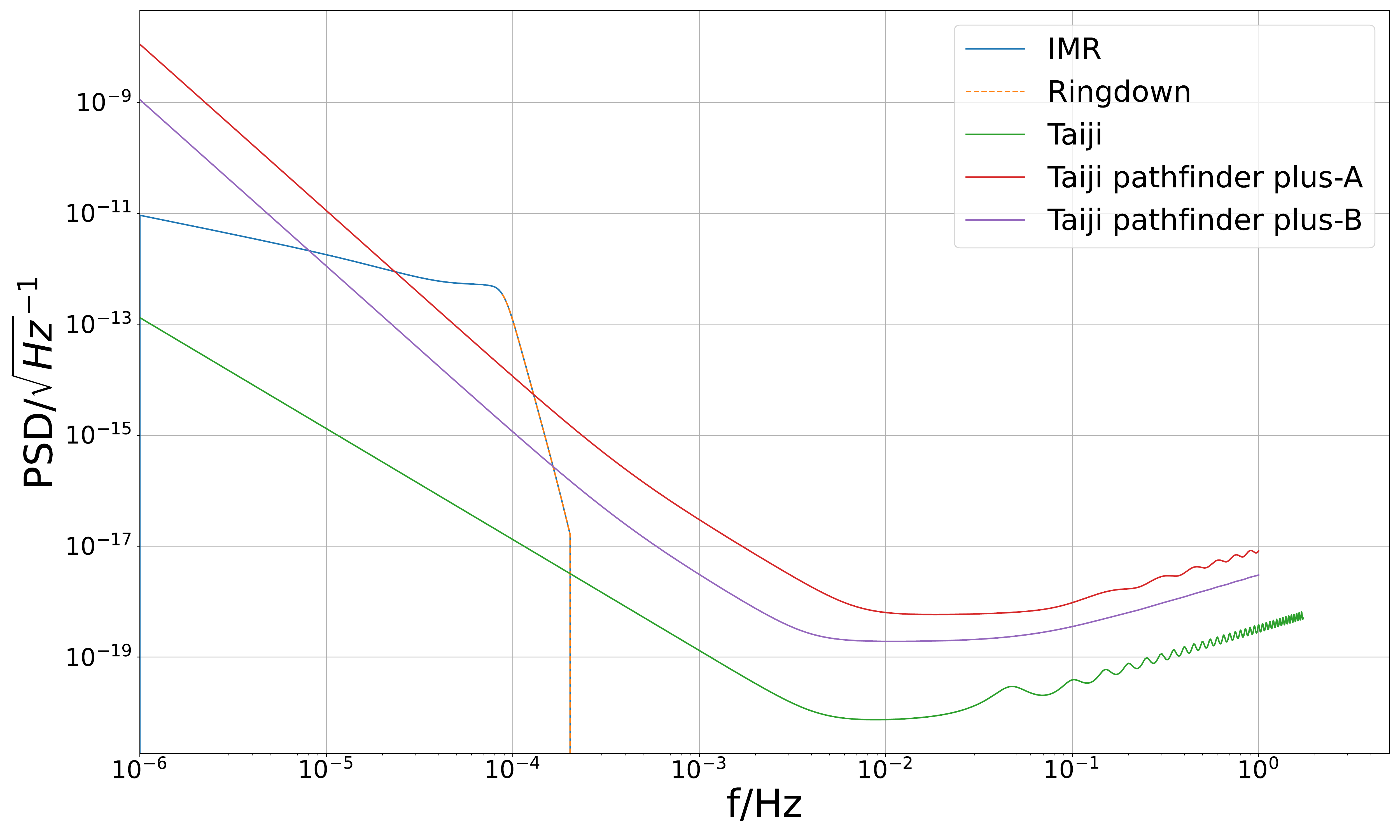}
\caption{\label{rdf} The PSDs and the equivalent for source amplitudes $\sqrt{S_n(f)}=2\sqrt{f}|\tilde{h}(f)|$ \cite{PSD}. The blue line is the amplitude equivalent of the inspiral-merger-ringdown frequency-domain signal, the ringdown signal corresponds to the orange line. The three other curves correspond to the PSD for pathfinder plus-A, pathfinder plus-B, and Taiji.}
\end{figure}

Fig. \ref{td_seobnre} shows the waveform for a circular orbit with eccentricity $e=0$ and for an orbit with high eccentricity, where $e=0.9$. We observe that it is sufficient to approximate the waveform of J1430+2303 by a circular orbital waveform for the last stage of the binary BH evolution. 
We consider here 10 days before the merger, whose corresponding starting frequency is about $10^{-5}$ Hz. 
In Fig. \ref{rdf}, we notice that the GW signal emitted during the final stage of the merger enters the detection band of Taiji plus-B. In the case of the final Taiji configuration, it seems preferable to use a waveform with a non-zero eccentricity to conduct the analysis, however for simplicity we will consider a circular orbital waveform with a starting frequency of $10^{-6}$ Hz. 
We consider 8 parameters to describe the gravitational source: the masses of the primary and secondary SMBHs denoted by $m_1$ and $m_2$ respectively, the primary SMBH spin $\chi_1$, the polar and azimuthal angles of the source's angular momentum in the ecliptic (detector) frame $\theta$ and $\phi$ ($\theta_L$ and $\phi_L$) respectively, and the luminosity distance $D_L$. The conversion from $\phi_L$ and $\theta_L$ to the polarization phase $\psi$ and the source inclination $\iota$ are given in Eq. (\ref{iota}). The spins of the binary have not yet been determined. Without loss of generality, we only consider a non-zero spin for the primary black hole with a dimensionless spin $\chi_1=0.4$ (and so $\chi_2=0$). The source inclination $\iota$ is $0.3\pi$ rad and the polarization phase is $\psi=0.3\pi$ rad. 
\begin{subequations}
\label{iota}
\begin{align}
\cos\iota =& \cos\theta_L\sin(\pi/2-\theta)\notag\\&+\sin\theta_L\cos(\pi/2-\theta)\cos(\phi_L-\phi)
\\
\tan\psi =& \frac{\cos\theta_L+\cos\iota\sin(\pi/2-\theta)}{\cos(\pi/2-\theta)\sin\theta_L\sin(\phi_L-\phi)}
\end{align}
\end{subequations}

\begin{figure*}[ht]
\centering
\includegraphics[width=1\textwidth]{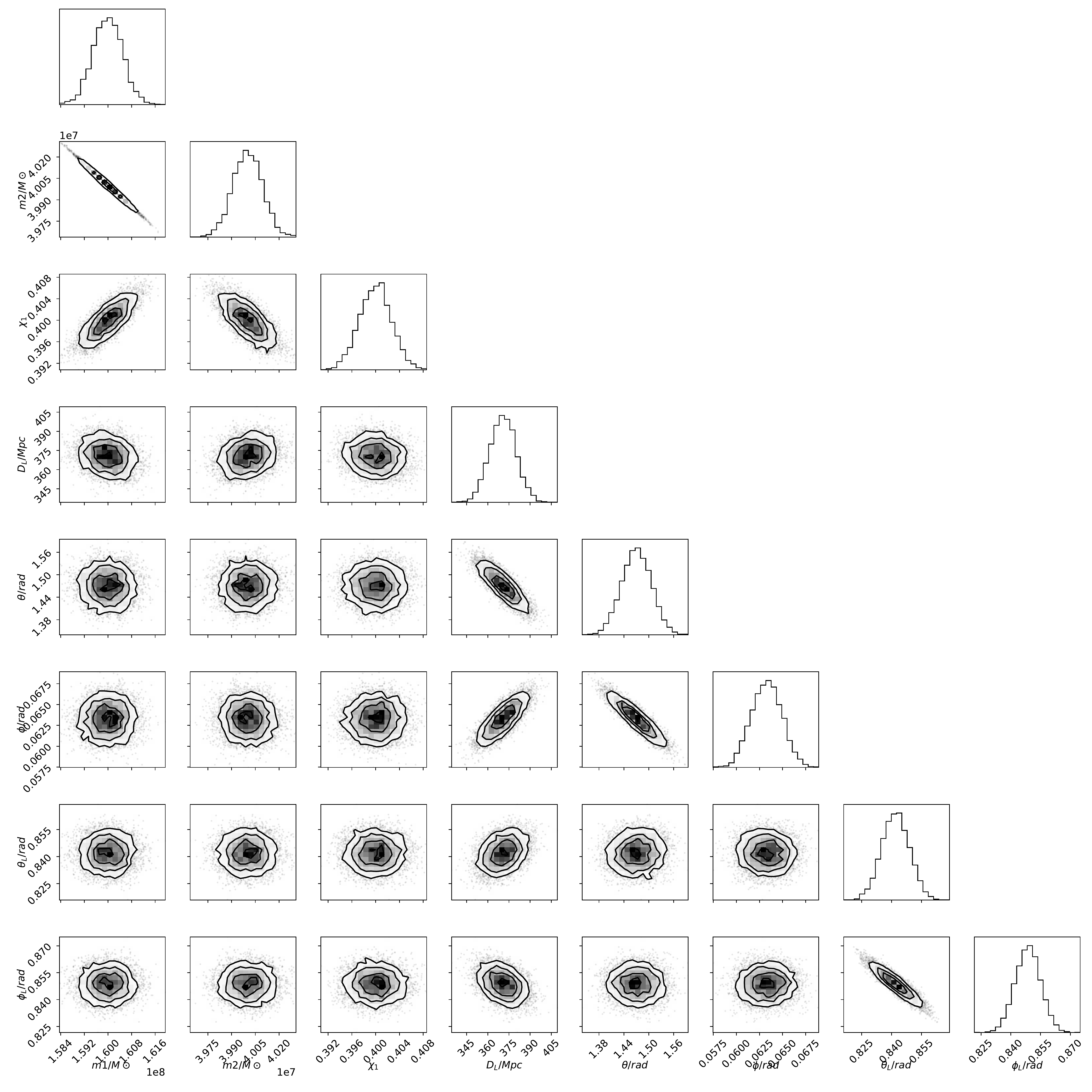}
\caption{\label{para}The probability distribution of parameters $m_1$, $m_2$, $\chi_1$, $\theta$, $\phi$, $\theta_L$, $\phi_L$ of the SMBHB observed by pathfinder plus-B.}
\end{figure*}

For the three different Taiji configurations, due to the uncertainties on the spins and inclination of the binary, the SNR values given by Eq. (\ref{SNR}) vary greatly. For example, the SNR measured in pathfinder plus-B ranges from 70 to 800. Here, the parameters we assumed are more conservative, the SNRs are about 13, 130, and 16000 corresponding to pathfinder plus-A, pathfinder plus-B, and Taiji. For a signal detected with an SNR beyond 10, the parameter uncertainties can be approximated as the square root of the diagonal elements of the inverse of $\Gamma_{ij}$, i.e., $\Delta \lambda_i \approx \sqrt{(\Gamma^{-1})_{ii}}$. We then use Fisher information matrix (FIM) method to perform the parameter estimation. The FIM for a frequency domain GW signal $\tilde{h}(f)$ parameterized by $\boldsymbol{\lambda}$ is given in \cite{cutler94},
\begin{align}
\Gamma_{ij} = \left<\frac{\partial \tilde{h}}{\partial \lambda_i}|\frac{\partial \tilde{h}}{\partial \lambda_j}\right>
\end{align}
with
\begin{align}
\label{snreq}
\left<\tilde{h}_1|\tilde{h}_2\right>=2{\rm Re}\int^{\infty}_0\frac{\tilde{h}_1^*(f)\tilde{h}_2+\tilde{h}_1(f)\tilde{h}_2^*(f)}{S_n(f)}df
\end{align}
\begin{align}
\label{SNR}
{\rm SNR}^2=4{\rm Re}\int^{\infty}_0\frac{\tilde{h}(f)\tilde{h}^*(f)}{S_n(f)}df
\end{align}
where $\tilde{h}(f)$ is the frequency-domain waveform, $S_n(f)$ is the noise power spectral density (PSD) of the detector, $\boldsymbol{\lambda}$ is the waveform parameter , and $S_n$ is the noise PSD of the detector. The results are presented in Table \ref{para_t}. We notice that the luminosity distance $D_L$ can be well measured by the pathfinder plus-B with an error of about $\Delta \lambda_i/\lambda_i = 5\% $. 
We also compute the probability distribution $\mathcal{L}({\boldsymbol\lambda}) \propto e^{-\frac{1}{2}\Gamma_{ij}\Delta\lambda_i\Delta\lambda_j}$ \cite{cutler94,Babak2017PhRvD..95b4010B} for pathfinder plus-B. The results are shown in Fig. \ref{para}. 

%
%
%

%
\begin{table*}[t]
\footnotesize
\caption{\label{para_t}Parameter estimation of the source. The results for Taiji are listed just as reference.}
\tabcolsep 11pt 
\begin{tabular*}{\textwidth}{c|c|c|c|c|c|c|c|c}
\hline\hline
 &$\Delta m_1/m_1$&$\Delta m_2/m_2$&$\Delta \chi_1/\chi_1$&$\Delta D_L/D_L$&$\Delta \theta/\theta$&$\Delta \phi/\phi$&$\Delta \theta_L/\theta_L$&$\Delta \phi_L/\phi_L$\\
 \hline
pathfinder plus-A&$6.5\times10^{-2}$&$5\times10^{-2}$&$1.5\times10^{-1}$&$5\times10^{-1}$&$5\times10^{-1}$&$5\times10^{-1}$&$2\times10^{-1}$&$2\times10^{-1}$\\
 \hline
pathfinder plus-B&$6\times10^{-3}$&$5\times10^{-3}$&$1\times10^{-2}$&$5\times10^{-2}$&$5\times10^{-2}$&$5\times10^{-2}$&$2\times10^{-2}$&$1.5\times10^{-2}$\\
\hline
Taiji&$6\times10^{-5}$&$4.5\times10^{-5}$&$1\times10^{-4}$&$1.5\times10^{-4}$&$6\times10^{-5}$&$6\times10^{-5}$&$2\times10^{-5}$&$2\times10^{-5}$\\
\hline\hline
\end{tabular*}
\end{table*}
\section{Key scientific goals with an imminant merging SMBHB}

In the final stage of the evolution of J1430+2303, this SMBHB will emit strong GW radiation. Although the signal is concentrated in a lower frequency band due to the enormous mass of the source, the expected signal amplitude is high enough to allow a detection by Taiji pathfinder plus-A, and the more sensitive detectors: Taiji pathfinder plus-B and the final configuration of Taiji. By studying the signal, we can derive a rich set of physical information. The discussion of potential science by J1430+2303 is divided into seven subsections below. In this part we will use the geometric units $G=c=1$.

Subsection ``Standard siren" briefly introduces the Hubble law and discusses the possibility of testing using J1430+2303. 

In GR, a Kerr BH is fully characterized by only two parameters: mass and spin. The ringdown signal from the merger remnant is described by the mass and the spin. Therefore, detecting this signal is a way to test the nature of black holes. We discuss the possibility of detecting such a signal using pathfinder plus-A and pathfinder plus-B in subsection ``Ringdown".

Subsection ``GW echoes" focus on exotic compact objects (ECOs), which are expected to produce a GW signal echo. The detection of echoes will provide the first evidence of the existence of ECOs and allow us to explore new physics. We show that Taiji pathfinder plus-B will be able to detect such a signal or place constraints on the existence of ECOs in the absence of a detection. 

The next subsection, ``Black hole area", discusses how pathfinder plus-B can be used to test Hawking's area theorem (or the second law of BH mechanics) with J1430+2303.

Subsection ``Gravitational recoil" is dedicated to the gravitational recoil, caused by the anisotropic emission of gravitational radiation. We investigate the possibility of using Taiji pathfinder plus-B to detect this astrophysical phenomenon. 

In subsection "GW dispersion", we compute the dispersion of the GW signal of J1430+2303. Based on a detection, we study the ability of the three types of detectors to place a constraint on the $\lambda_g$ parameter which can be used to test GR.

Finally, the last subsection ``ppE and polarization tests" presents how the accuracy of ppE parameters and polarizations can be estimated with the detection of J1430+2303 using Taiji plus-B and the final configuration of Taiji.

Note that Taiji may not catch this source on time, the results for Taiji are just listed as references.

\subsection{Standard siren}
The strength of the GW signal is proportional to the source distance. Generally, because the detected frequency of the GW signal is one after the red-shift, the mass and distance we measured are also red-shifted. Consequently the signal does not provide a direct measure of the redshift $z$ , but a measure of the luminosity distance $D_L$.

In the low redshift range, the redshift $z$ and the luminosity distance $D_L$ satisfy the Hubble law:
\begin{align}
\label{hubble law}
v_{\rm H}=H_0D_L
\end{align}
where $H_0$ is the Hubble constant and $v_{\rm H}$ is the receding velocity. The red-shift $z$ and the receding velocity $v_{\rm H}$ satisfy the relation $z=v_{\rm H}/c$. Therefore, a limit on the Hubble constant $H_0$ can be derived by measuring the receding velocity $v_{\rm H}$ and the luminosity distance $D_L$.

We know the host galaxy of the SMBHB, J1430+2303, so we have access to a measure of the redshift. 
We show in Table \ref{para_t} that both detectors Taiji plus-B and the final Taiji configuration can accurately determine the luminosity distance $D_L$. Thus we can obtain a measure of the Hubble constant. 

If we compare the measurement of the luminosity distance $D_L$ by ground-based detectors, space-based detectors have a great advantage, namely the much longer duration of the target signal and a long orbit around the sun to estimate $D_L$ at different angles. This will greatly improve the measurement of $D_L$. The relative error on the $D_L$ measurement for the third observing run (O3) \cite{O3} is about $30\%$. For pathfinder plus-A, the relative error would be around $50\%$; but for pathfinder plus-B, the relative error will reduce to only 
$5\%$. Hence, pathfinder plus-B has the ability to precisely measure the Hubble constant with $\Delta H_0 \approx3\rm km s^{-1}Mpc^{-1}$ in the case of a signal detection from the J1430+2303 source. As a reference, the final Taiji will give $\Delta H_0\approx1\times10^{-2} \rm km s^{-1}Mpc^{-1}$.

\subsection{Ringdown}

The part of the GW signal related to the oscillations of a single remnant BH is known as the "ringdown" phase, as the perturbed BH rings like a bell. From the no hair theorem, we know that a Kerr BH is fully described by only its mass and spin. The ringdown signal can be expressed as a superposition of quasi-normal modes (QNM) with a complex frequency, where the real part represents the oscillation frequency and the imaginary part gives the inverse damping time of the mode \cite{PhysRevD.73.064030, QNM1999}. In GR, the oscillation frequency and the damping time are completely determined by the mass and spin of the remnant BH. The measurement of QNMs can be used for studying strong gravity, characterizing the remnant BH, and testing GR \cite{PhysRevD.73.064030, QNM1989, PhysRevD.46.5236, PhysRevD.57.4535, QNM2004}. 

We have no information on the spin of the initial BHs, we set the dimensionless spin of the primary SMBH to $\chi_1=0.4$ and no spin for the second BH ($\chi_2=0$). We calculate the final mass $M_{f}$ and the final spin $ \chi_{ f}$ of the remnant BH using the fitting formulas given in \cite{PhysRevD.82.064016}.

\begin{figure}
\centering
\includegraphics[width=0.49\textwidth]{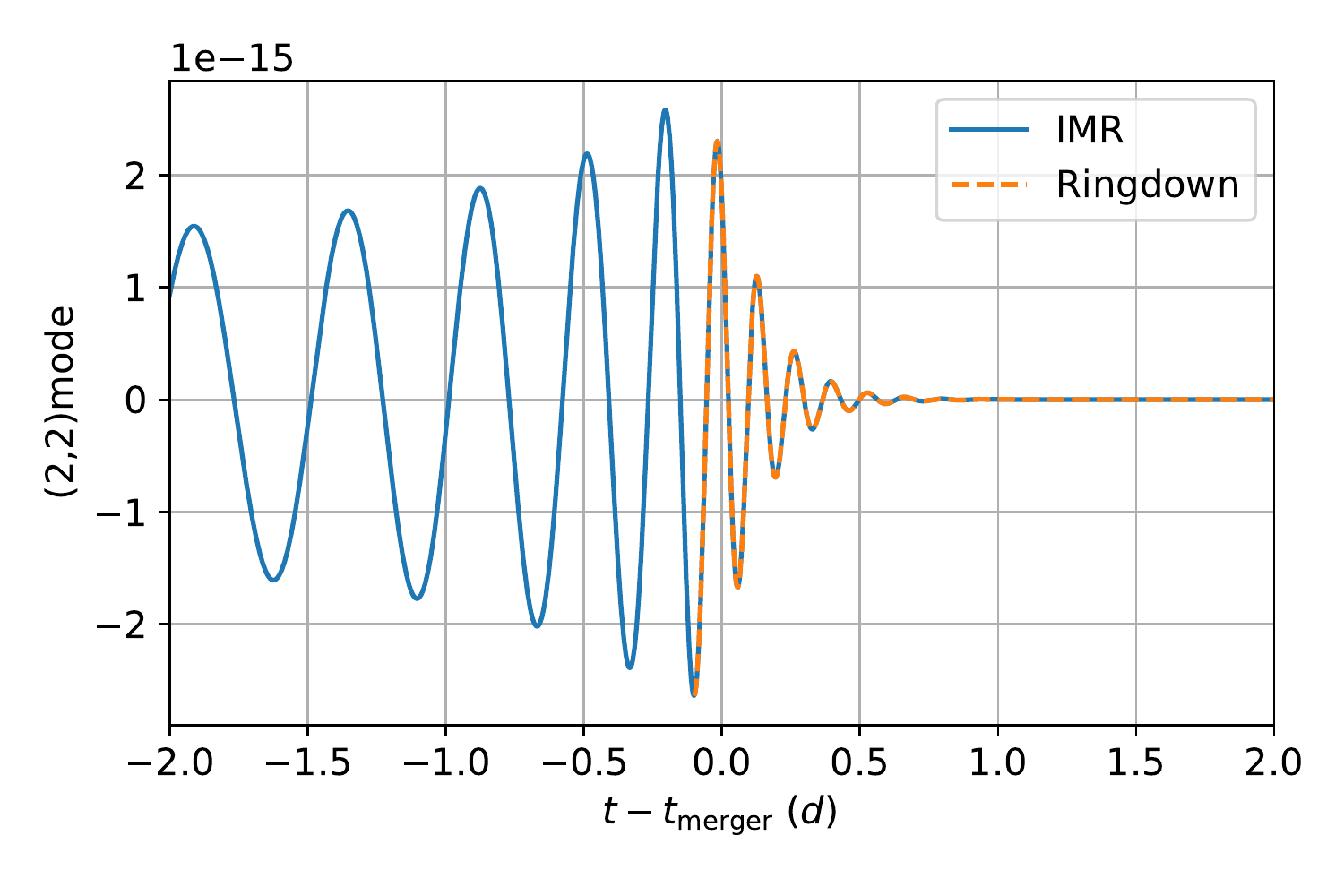}
\caption{\label{rd}The IMR (in blue) and ringdown (in orange) waveforms for J1430+2303. The ringdown waveform is obtained by cutting the IMR waveform at $t_{\rm peak}$. The difference of time, $\Delta t$, between $t_{\rm peak}$ and $t_{\rm merger}$ is a time difference between the Teukolsky $(2, 2)$ mode amplitude peak and the EOB orbital frequency peak \cite{PhysRevD.85.024046}.}
\end{figure}

Fig. \ref{rd} shows the time domain IMR and ringdown waveforms which are generated with ``SEOBNRv4" method. Fig. \ref{rdf} presents the PSD. We use here the simplest way to obtain the ringdown signal, by truncating the IMR signal at the time $t_{\rm peak}$ of maximum amplitude.

\begin{table}
\caption{\label{rdsnr}The SNR of the ringdown signal.}
\centering
\begin{tabular}{c|c|c|c}
\hline\hline
 &pathfinder plus-A&pathfinder plus-B&Taiji\\
 \hline
 SNR&6&62&5800
 \\
\hline\hline
\end{tabular}
\end{table}

\begin{figure}
\centering
\includegraphics[width=0.46\textwidth]{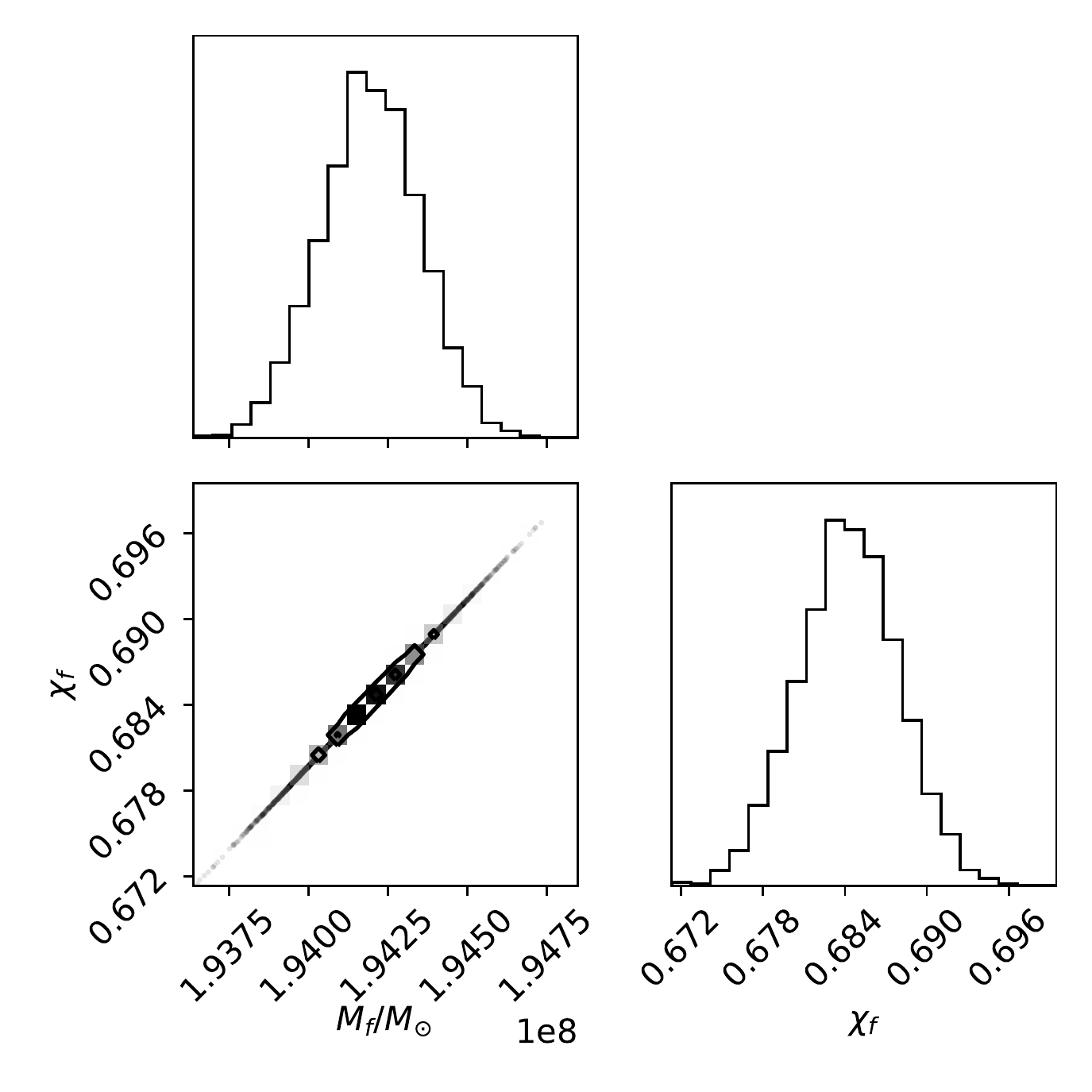}
\caption{\label{rdp} Probability distributions for the final mass $M_{ f}$ an for the final spin $ \chi_{ f}$ of the SMBHB observed by pathfinder plus-B.}
\end{figure}

As we can notice from Table \ref{rdsnr}, the ringdown signal can be detected with high SNR, thus the final mass and spin of the remnant can be measured accurately and the no-hair theorem will be tested at a precise level. To quantify the accuracy with which the parameters $ \chi_{ f}$ and $M_{ f}$ can be measured, we use the FIM method to perform parameter estimation. The results are presented in table \ref{rdpt} for the three configurations of the Taiji detector. In addition, the probability distributions $\mathcal{L}({\boldsymbol\lambda}) \propto e^{-\frac{1}{2}\Gamma_{ij}\Delta\lambda_i\Delta\lambda_j}$ \cite{cutler94, Babak2017PhRvD..95b4010B} for pathfinder plus-B are shown in Fig. \ref{rdp}. The final mass and spin precision measurement in O3 are around $10^{-2}$\cite{O3}.

\begin{table}
\caption{\label{rdpt}Parameter estimation for $ \chi_{ f}$ and $M_{f}$.}
\centering
\begin{tabular}{c|c|c}
\hline\hline
 &pathfinder plus-B&Taiji\\
 \hline
 $\Delta M_f/M_f$&$8\times10^{-4}$&$9\times10^{-6}$
 \\
 \hline
  $\Delta \chi_f/\chi_f$&$5\times10^{-3}$&$6\times10^{-5}$
 \\
\hline\hline
\end{tabular}
\end{table}

\subsection{GW echoes}

For BHs in GR, there is a natural boundary condition that requires only pure ingoing waves at the BH surface due to the horizon property. Reference \cite{PhysRevD.98.124009,cardoso19} indicated that the equations of state of exotic matter, phase transitions or quantum gravity allow the existence of ECOs such as boson stars, fluid stars, gravastars, bubbles, fuzzy balls, superspinars, etc. These objects are as compact as BHs but have no horizon. Because astrophysical processes are usually insensitive to the spacetime geometry near the horizon, highly compact ECOs behave very similarly to BHs. GW echoes are produced by the reflection of GW waves between the near horizon membrane barrier and the angular momentum barrier. The delay time $\Delta t_{\rm echo}$ between each echo is given by the travel time of the signal between these two barriers.

For a Kerr black hole, $\Delta t_{\rm echo}$ is 
\begin{equation}
    \begin{aligned}
    \Delta t_{\rm echo}=&2\times\int^{r_{\rm max}}_{r_++\Delta r}\frac{r^2+a^2M^2}{r^2-2Mr+a^2M^2}dr\\=&2r_{\rm max}-2r_+-2\Delta r+2\frac{r^2_++a^2M^2}{r_+-r_-}\ln(\frac{r_{\rm max}-r_+}{\Delta r})\\&-2\frac{r^2_-+a^2M^2}{r_+-r_-}\ln(\frac{r_{\rm max}-r_-}{r_+-r_-\Delta r})
    \end{aligned}\label{deltat}
\end{equation}
where $M$ and $a$ are the final mass $M_f$ and final spin $\chi_f$, $r_{\pm}=M(1\pm\sqrt{1-a^2})$, $\Delta r$ is the coordinate distance of surface (or membrane) and would-be horizon and $r_{\rm max}$ is the peak position of the angular momentum barrier which is given by the root of a polynomial equation, Eq. (2.18) in \cite{PhysRevD.88.044047}.

The membrane lies outside the horizon, at a Planck proper distance \cite{surface1983,surface1985,surface2009} given by:
\begin{align}
&\int^{r_++\Delta r}_{r_+}\sqrt{g_{rr}}dr|_{\theta=0}\sim l_p\approx1.62\times10^{-33}{\rm cm}.
\end{align}

The radial location of the membrane is then given by:
\begin{align}
&\Delta r|_{\theta=0}=\frac{\sqrt{1-a^2}l^2_p}{4M(1+\sqrt{1-a^2})}.
\end{align}
Here we used a sample time-domain echo template $\mathcal{M}_{\rm TE,I}(t)$ \cite{PhysRevD.96.082004} with five free parameters $A$, $\gamma$, $t_0$, $t_{\rm echo}$ and $\Delta t_{\rm echo}$ given by
\begin{subequations}
    \begin{align}
\mathcal{M}_{\rm TE,I}(t)\equiv& A\sum\limits_{n=0}^{\infty}(-1)^{n+1}\gamma^n\notag\\&\times\mathcal{M}_{\rm T,I}(t+t_{\rm merger}-t_{\rm echo}-n\Delta t_{\rm echo},t_0)\\\mathcal{M}_{\rm T,I}(t)\equiv&\Theta(t,t_0)\mathcal{M}_{\rm I}(t)\notag\\\equiv&\frac{1}{2}\{1+{\rm tanh}[\frac{1}{2}\omega_{\rm I}(t)(t-t_{\rm merger}-t_0])\}\mathcal{M}_{\rm I}(t)
    \end{align}
\end{subequations}
where $A$ is the overall amplitude of the echo template, $\gamma$ is the damping factor between successive echoes, $t_{\rm merger}$ is the time of the merger, $t_{\rm echo}$ is the time of the first echo, and the free parameter $t_0\in[-0.1,0]{\Delta t}_{\rm echo}$ is used to cut the last part of the original waveform to produce the echo waveform, $\mathcal{M}_{\rm T, I}(t)$ is a smooth activation of the GW template, $\omega_{\rm I}(t)$ denotes the angular frequency evolution of the frequency-domain waveform as a function of time, and $M_{\rm I}(t)$ is the IMR waveform which is generated by the "SEOBNRv4" approximant. The smooth activation $\Theta(t, t0)$ essentially selects the ringdown signal, which is the part of the waveform that one might expect to see in echo signal.

\begin{figure}
\centering
\includegraphics[width=0.48\textwidth]{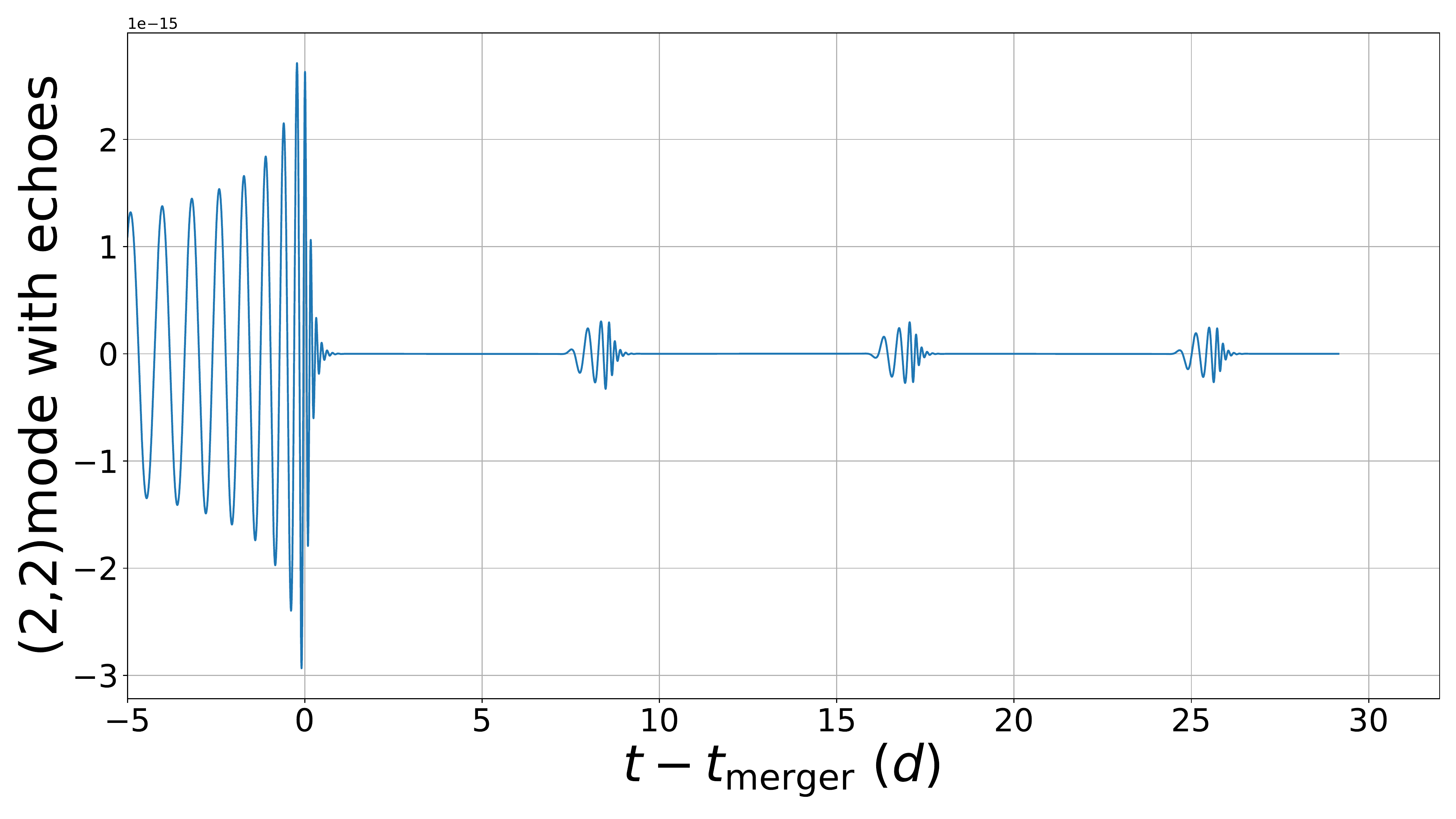}
\caption{\label{td_e} Time-domain waveform and the first three of the resulting echoes. The echo waveform parameters are choosen to be: $\Delta t_{\rm echo}=10.15652(d)$, $\gamma=0.9$, $A=0.124$, $t_0/\Delta t_{\rm echo}=-0.1$ and $(t_{\rm echo}-t_{\rm merger})/\Delta t_{\rm echo}=1$. }
\end{figure}

\begin{figure}
\centering
\includegraphics[width=0.48\textwidth]{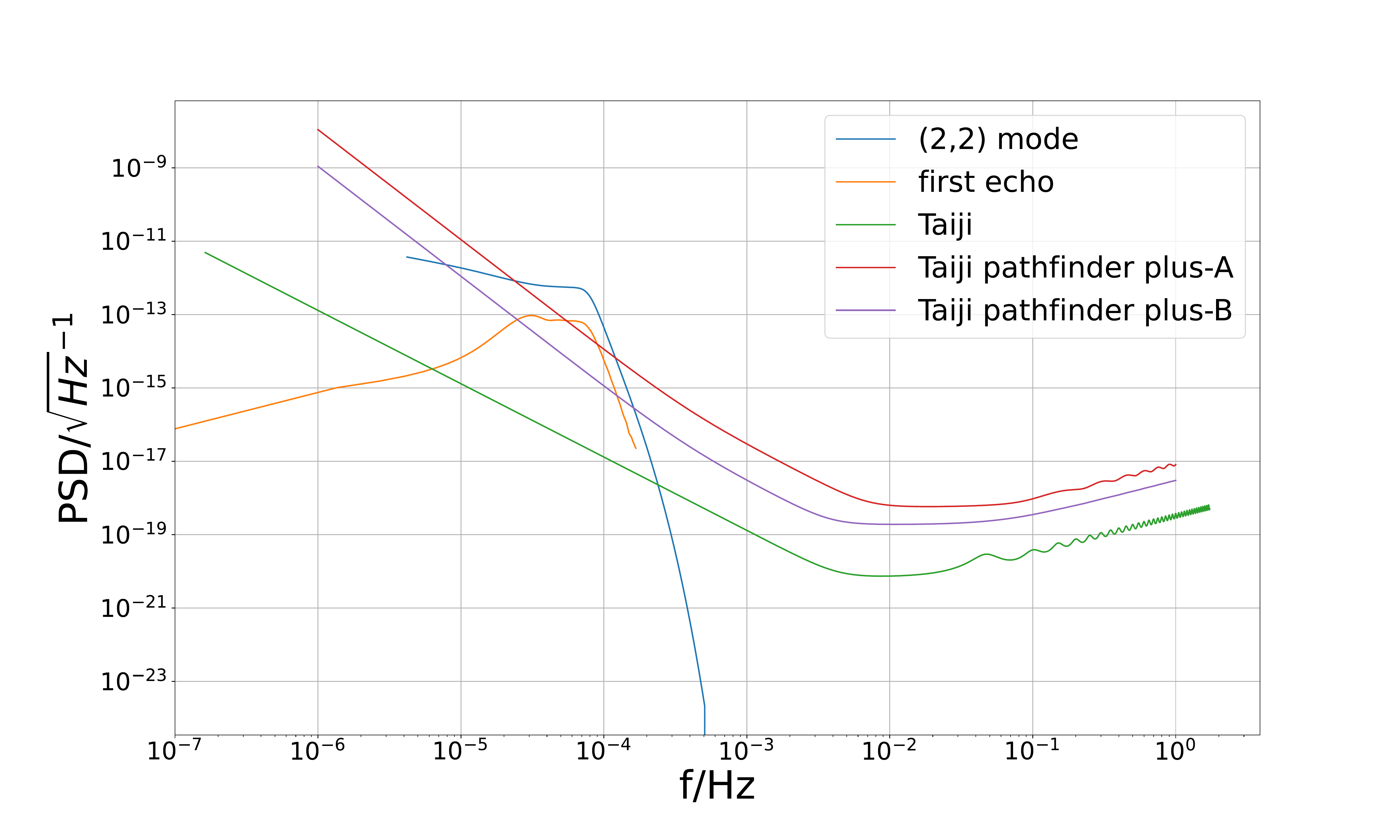}
\caption{\label{fd_e}The PSD, where equivalent for source amplitude $\sqrt{S_n(f)}$ is computed by the Fourier transformation of the waveform in Fig. \ref{td_e}.}
\end{figure}

The free parameter $\Delta t_{\rm echo}=10.15652$ days (d) is computed using Eq. (\ref{deltat}) and we set the four remaining free parameters at $\gamma=0.9$, $A=0.124$, $t_0/\Delta t_{\rm echo}=-0.1$ and $(t_{\rm echo}-t_{\rm merger})/\Delta t_{\rm echo}=1$ considering the best fit values of Ref. \cite{PhysRevD.96.082004}. Here, Fig. \ref{td_e} and \ref{fd_e} show the time-domain waveform and PSD. Table \ref{echosnr} shows the SNRs measured for the first echo using different values of $\gamma$. The GW echo signal is strong enough to be detected by pathfinder plus-B and Taiji.   

In addition, we use the FIM method to estimate the damping factor between successive echoes, $\gamma$. The results are presented in Table \ref{echop}, we found that pathfinder plus-B and Taiji could measure the damping factor, $\gamma$, with respectively a relative error of $10^{-2}$ and $10^{-4}$ order.

\begin{table}
\caption{\label{echosnr} First echo SNR for different  values of $\gamma$.}
\centering
\begin{tabular}{c|c|c}
\hline\hline
&pathfinder plus-B&Taiji\\
 \hline
 $\gamma=0.6$&11&1289
 \\
 \hline
$\gamma=0.9$&16&1935
 \\
\hline\hline
\end{tabular}
\end{table}
\begin{table}
\caption{\label{echop}Parameter estimation for $\gamma$ using the first and second echo GW signal.}
\centering
\begin{tabular}{c|c|c}
\hline\hline
&pathfinder plus-B&Taiji\\
 \hline
 $\Delta\gamma/\gamma$&$6\times10^{-2}$&$5\times10^{-4}$
 \\
\hline\hline
\end{tabular}
\end{table}

\subsection{Black hole area}
Hawking’s area theorem \cite{Hawking_area} tells us that the total horizon area of classical BHs cannot decrease over time. According to this, the combined horizon area of the two progenitor BHs must not exceed that of the remnant. Thus, by computing the area of the two BHs before the merger and the remnant area, we can test this theorem.

The horizon area, $\mathcal{A}$, of a Kerr BH with mass $m$ and spin $\chi$ is given by \cite{BHA_prl}
\begin{equation}
 \mathcal{A}(m,\chi)=8\pi m^2(1+\sqrt{1-\chi^2}) \label{A_eq}.
\end{equation}
From Eq. (\ref{A_eq}), we notice that to measure the BH area $\mathcal{A}$ we need to estimate precisely two parameters: the mass and the spin, before and after the merger. Assuming that we have no preliminary information on the spin of these two SMBHs, we consider that $\chi_1$ varies from $0$ to $0.98$ and we set the spin of the secondary BH, $\chi_2$, equal to zero. The area $A_0$ of the two progenitor BHs is $8\pi m_1^2(1+\sqrt{1-{ \chi_{1}}^2})+ 16\pi m_2^2$, and the remnant BH area is $8\pi m_{\rm f}^2(1+\sqrt{1-{\chi_{\rm f}}^2})$. Here, $m_{\rm f}$, $ \chi_{\rm f}$ are the remnant mass and spin respectively, which can be derived from the fitting formula given in \cite{final_mass_spin_prd,final_mass_spin_prd_apj}.

\begin{figure}
\centering
\includegraphics[width=0.49\textwidth]{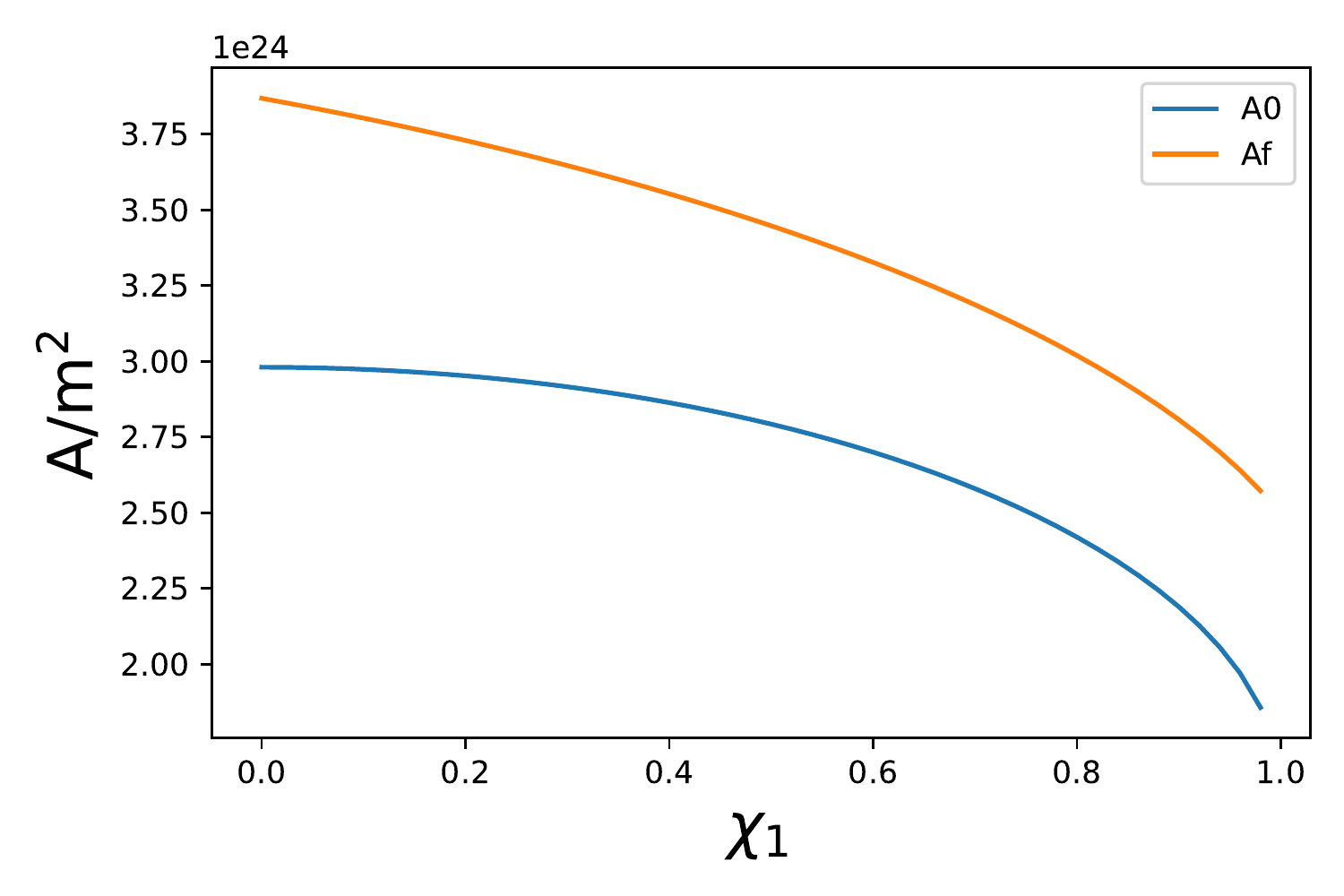}
\caption{\label{Area} $A_{0}$ represents the sum of the intial two BH areas before the merger (in blue) and $A_{f}$ is the remnant BH area (in orange). The BH area increase after the merger, as stated by the Hawking's area theorem.}
\end{figure}

Fig. \ref{Area} shows the variation of the BH area as a function of the spin $\chi_1$. The orange line represents the area, $A_{f}$, of the remnant based on the current theoretical prediction for $m_{\rm f}$ and $ \chi_{\rm f}$, which obeys to the Hawking’s area theorem. Due to the high precision of mass and spin measurements of the two components ($10^{-3}$ and $10^{-2}$ respectively by pathfinder plus-B from Table.\ref{para_t}) and the final BH ($10^{-4}$ and $10^{-3}$ respectively from Table.\ref{rdpt}), we can test Hawking’s area theorem 
very accurately comparing to the current result by Advanced LIGO \cite{BHA_prl}. 

\subsection{Gravitational recoil}
The emission of gravitational radiation drives the orbital evolution of BHBs towards an eventual merger. By conservation of momentum, there is a recoil or kick velocity $\bf v_k$ \cite{kick_prl}  of the final remnant. The velocity $v_k$ can have a value between $170 \rm km/s$ \cite{PhysRevLett.98.091101} and $5000\rm km/s$ \cite{PhysRevLett.98.231101,PhysRevLett.98.231102,PhysRevLett.107.231102}. Recently, Ref. \cite{kick2022} shows evidence for large kick velocities around $1500 \rm km/s$ or at least $\rm 700 km/s$ (one-sided limit).

If the direction of the kick velocity is not perpendicular to the line of sight, the gravitational recoil will appear as a gradually increasing GW Doppler shift during the final cycles and the merger. Actually, it can be equivalent to rescale the binary’s total mass $M$ in the phase evolution from $M$ to $M (1 + \bf v_k \cdot\hat{n})$, where the unit vector $\bf\hat{n}$ denotes the direction of the line of sight from the observer to the source.  We also call this shift the Doppler mass shift. Here, we only consider the non-relativistic case.

The GW frequency $f$ has always the dimensionless form $Mf$ in the binary dynamics, where $M$ is the total mass of the source. This indicates that the frequency shift and the total mass $M$ are degenerate. The Doppler mass shift produced by gravitational recoil is different from the usual frequency shift, such an effect gradually accumulates during the last orbits and the merger, while the cosmological redshift will affect the whole GW signal. Therefore, we expect to detect this effect separately.

\begin{figure}
\centering
\includegraphics[width=0.49\textwidth]{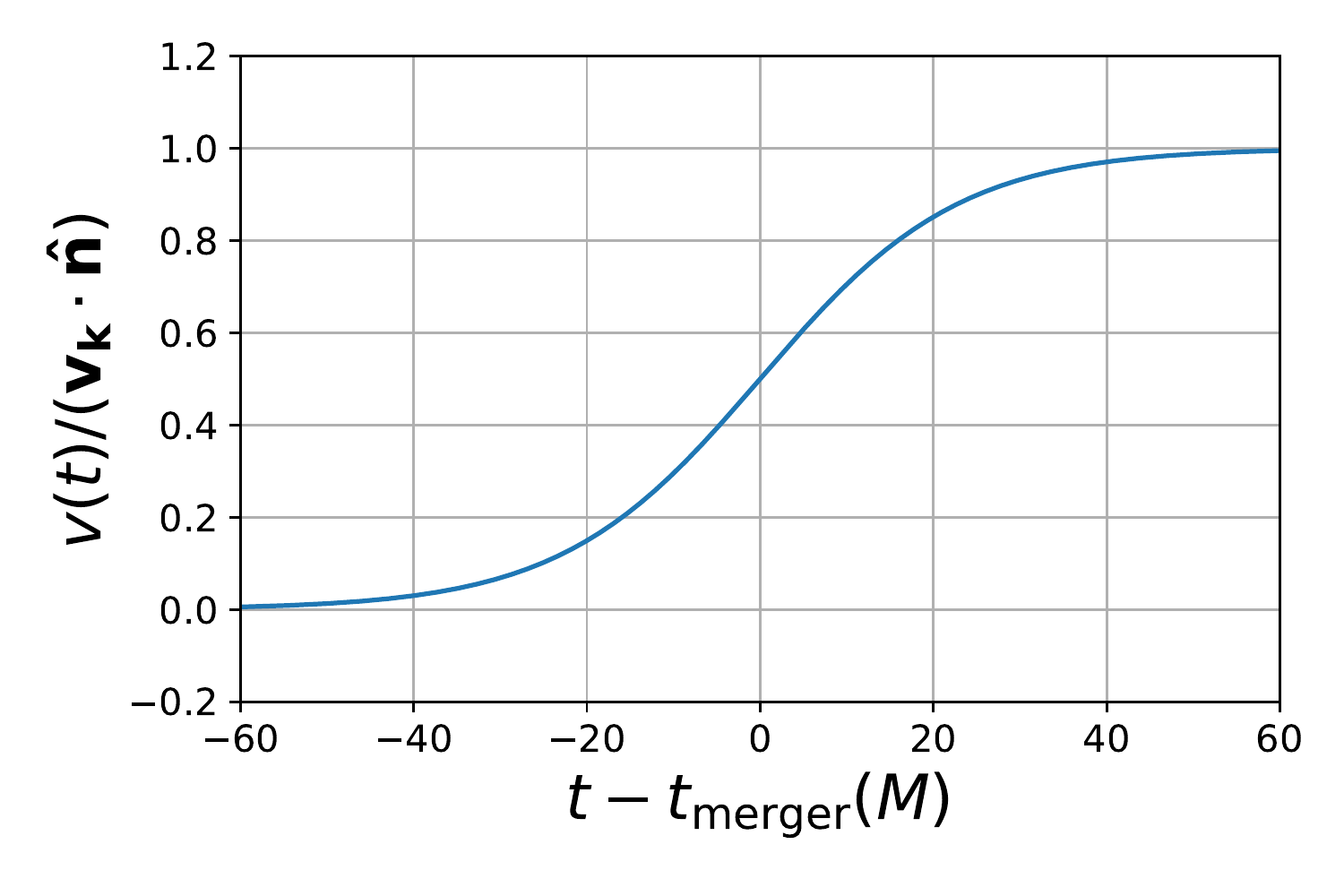}
\caption{\label{kick} kick velocity profile $v_k(t)$ for a duration time $\sigma = 20M$ and $\alpha_1/\alpha_0 = 0$.}
\end{figure}

Here we adopt a flexible model for the kick velocity profile and expand $\frac{d}{dt}v_k(t)$ which is given by \cite{kick_prl}
\begin{subequations}
    \begin{align}
        \frac{d}{dt}v_k(t)&={\bf V_k\cdot n}\frac{\sum_n\alpha_n\phi_n(t)}{\int^{\infty}_{-\infty}\sum_n\alpha_n\phi_n(t)dt}\\
        \phi_n(t)&=\frac{1}{\sigma \sqrt{2^nn!\sqrt{\pi}}}{\rm exp}(-\frac{(t-t_c)^2}{2\sigma^2})H_n(\frac{t-t_c}{\sigma})
    \end{align}
\end{subequations}
where $H_n$ are the Hermite polynomials, $t_c$ is the time of coalescence, $\sigma$ gives the duration of the recoil and the $\alpha_n$ weighs the various components so that the $\phi_n(t)$ constitute a complete basis in order to model all possible kick velocity profiles. We set $\sigma/M=10$ and $\alpha_1/\alpha_0=0$ (only one component, $\alpha_n = 0 $ when $n\geq1$) to construct the kick model. Fig. \ref{kick} presents the kick velocity profile model for $v_K(t)$.

\begin{figure}
\centering
\includegraphics[width=0.49\textwidth]{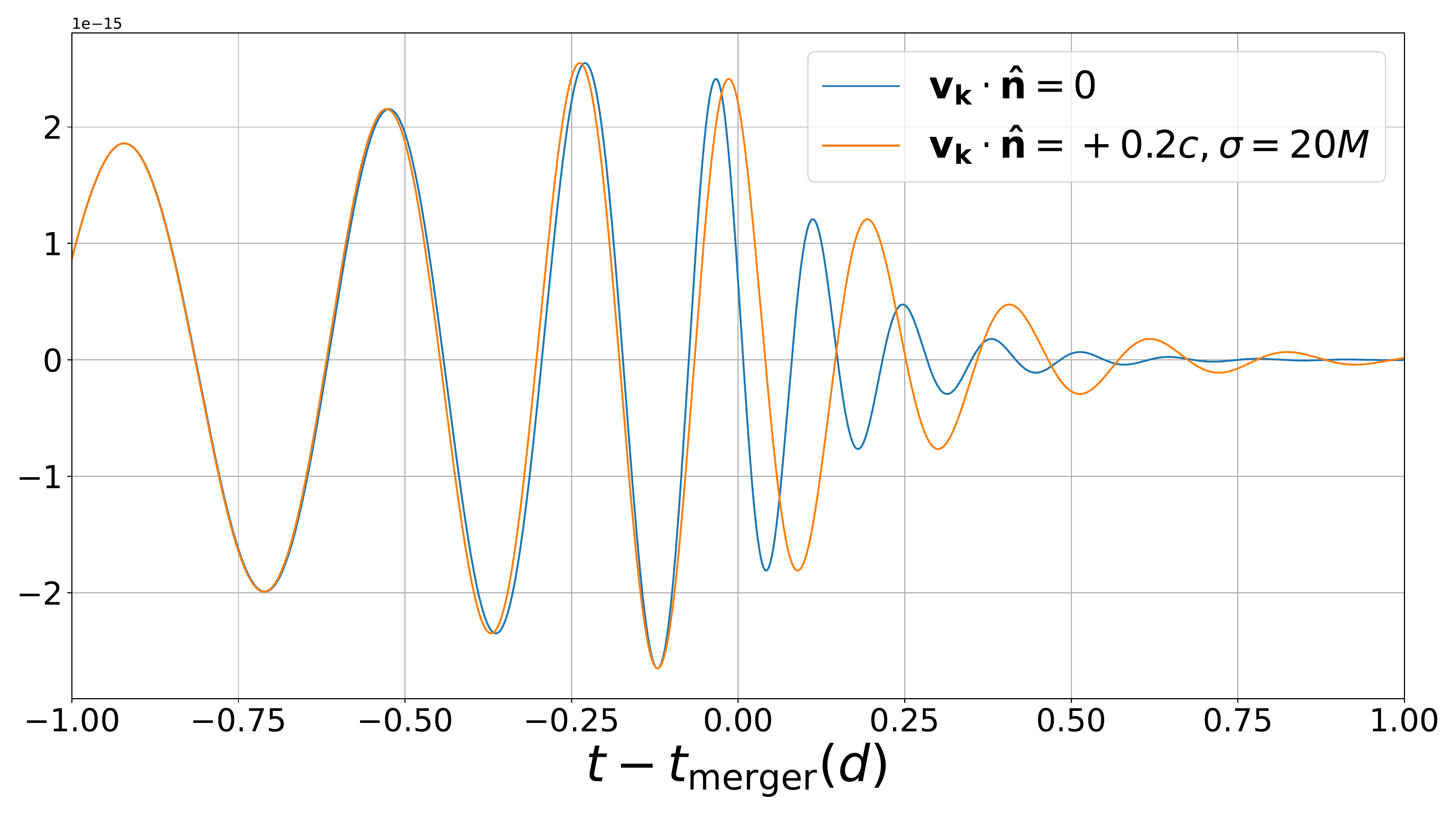}
\caption{\label{kgw}Time-domain GW waveforms. The blue line represents the waveform in the absence of a gravitational recoil, and the orange line shows the waveform with a kick velocity $v_k(t) =+0.2c$.}
\end{figure}

Fig. \ref{kgw} illustrates the difference between the waveform considering or not a gravitational recoil. We fixed the kick velocity at $v_k(t) =+0.2c$ and the duration of gravitational recoil at  $\sigma=20M$. The value of the kick velocity is deliberately chosen very high to clearly illustrate the difference. 

For the following estimation of $v_k$, we set the waveform parameters to $m_1=1.6\times10^8M_{\odot}$, $m_2=4\times10^7M_{\odot}$, $\chi_1=0.4$ for the mass, the inclination is $\iota=0.3\pi$ rad and the luminosity distance  is $D_L=370.7 \rm Mpc$. For the model part, the kick velocity is given by $v_k =+2\times10^{-4}c\approx60$km/s, the gravitational recoil duration is chosen to be $\sigma = 20M$, and $\alpha_1/\alpha_0 = 0$.

In order to estimate the detectability of $v_k$, we need to compute the match between the standard waveform and the waveform predicted in presence of a gravitational recoil,
\begin{equation}
\label{meq}
{\rm match}={\rm max}[\frac{\left<h_1|h_2\right>}{\sqrt{\left<h_1|h_1\right>\left<h_2|h_2\right>}}]
\end{equation}
where the ``max'' stands for the maximum match given by optimizating over a time shift.

As presented in reference \cite{0p995}, we set $0.995$ as the criterion value, and consider that if the match is smaller than $0.995$ we can distinguish the two waveforms. Our analysis found that for a detection using the sensitivity of Taiji pathfinder plus-B, if the kick velocity is $v_k \gtrsim 900$km/s this criterion is satisfied, and we can clearly discern the gravitational recoil in the signal.

\subsection{GW dispersion}
The major difference between GR and some modified gravity theories lies in the dispersion relation. In GR, the rest mass of gravitons $m_g$ must be equal to zero, while this is not necessarily the case in other theories. This affects the Lorentz invariance relation which could deviate from it expected value in such case. There are some methods to test Lorentz invariance, including GW observations \cite{PhysRevD.85.024041} and methods that are independent of the choice of cosmological models \cite{Zhang_2018}. For example, a Lorentz-violating graviton dispersion relation could have the following form \cite{PhysRevD.85.024041}
\begin{equation}
E^2=p^2c^2+m_g^2c^4+\mathbb{A}p^{\alpha}c^{\alpha}
\end{equation}
where $m_g$ is the mass of the graviton, $\alpha$ and $\mathbb{A}$ are the Lorentz-violating parameters that characterize the differences between GR and modified gravity theories. These two parameters are specific to each theorie, where we note that A has a dimension of $[\rm energy]^{2-\alpha}$.
The dispersion waveform in the frequency domain can be generalized as 
\begin{equation}
\tilde{h}(f) = \tilde{A}(f)e^{i\Psi(f)}
\end{equation}
with
\begin{equation}
\Psi(f) = \Psi_{\rm GR}(f)+\delta\Psi(f)
\end{equation}
where the dephasing $\delta\Psi(f)$ is caused by the propagation effects. Considering the case $\alpha\neq1$ (general case) or $\alpha=1$ (special case), the dephasing is different such that:
\begin{equation}
\delta\Psi(f) =\left\{
\begin{aligned}
-\beta u^{-1}&-\zeta_{\alpha\neq1} u^{\alpha-1},\quad \alpha\neq1 \\
-\beta u^{-1}&-\zeta_{\alpha=1} {\rm \ln}(u),\quad \alpha=1
\end{aligned}
\right.
\end{equation}
with the parameters
\begin{subequations}
\begin{align}
\beta&\equiv\frac{\pi^2D_0\mathcal{M}}{\lambda_g^2(1+Z)}\\
\zeta_{\alpha\neq1}&\equiv\frac{\pi^{2-\alpha}}{1-\alpha}\frac{D_{\alpha}}{\lambda_{\mathbb{A}}^{2-\alpha}}\frac{\mathcal{M}^{1-\alpha}}{(1+Z)^{1-\alpha}}\\
\zeta_{\alpha=1}&=\frac{\pi D_1}{\lambda_{\mathbb{A}}}
\end{align}
\end{subequations}
where the distance, $D_{\alpha}$, is defined by 
\begin{equation}
D_{\alpha}\equiv(\frac{1+Z}{a_0})^{1-\alpha}\int^{t_a}_{t_e}a(t)^{1-\alpha}dt
\end{equation}
where $\mathcal{M}=(m_1m_2)^{3/5}/(m_1+m_2)^{1/5}$ is the chirp mass, $Z$ is the cosmological redshift, $\lambda_g$ is the graviton Compton wavelength, and $\lambda_{\mathbb{A}}$ is a parameter which is defined in Eq. (13) of \cite{PhysRevD.85.024041}.

In addition, in the dark energy-matter dominated universe, $D_{\alpha}$ and $D_L$ are determined by the following equations

\begin{subequations}
\begin{align}
D_{\alpha}&=(\frac{1+Z}{H_0})^{1-\alpha}\int^{Z}_{0}\frac{(1+z')^{\alpha-2}dz'}{\sqrt{\Omega_M(1+z')^3+\Omega_{\Lambda}}}\\
D_{L}&=\frac{1+Z}{H_0}\int^{Z}_{0}\frac{dz'}{\sqrt{\Omega_M(1+z')^3+\Omega_{\Lambda}}}
\end{align}
\end{subequations}
where $H_0$ is the Hubble parameter and $\Omega_M$  and $\Omega_{\Lambda}$ are the matter and dark energy density parameters today. Here, we set $D_L = 370.7 \rm Mpc$.

To quantify the difference between the waveform predicted by GR and other gravity theories, we use Eq. (\ref{meq}) to derive the maximum match between these two waveforms.

\begin{figure}
\centering
\includegraphics[width=0.48\textwidth]{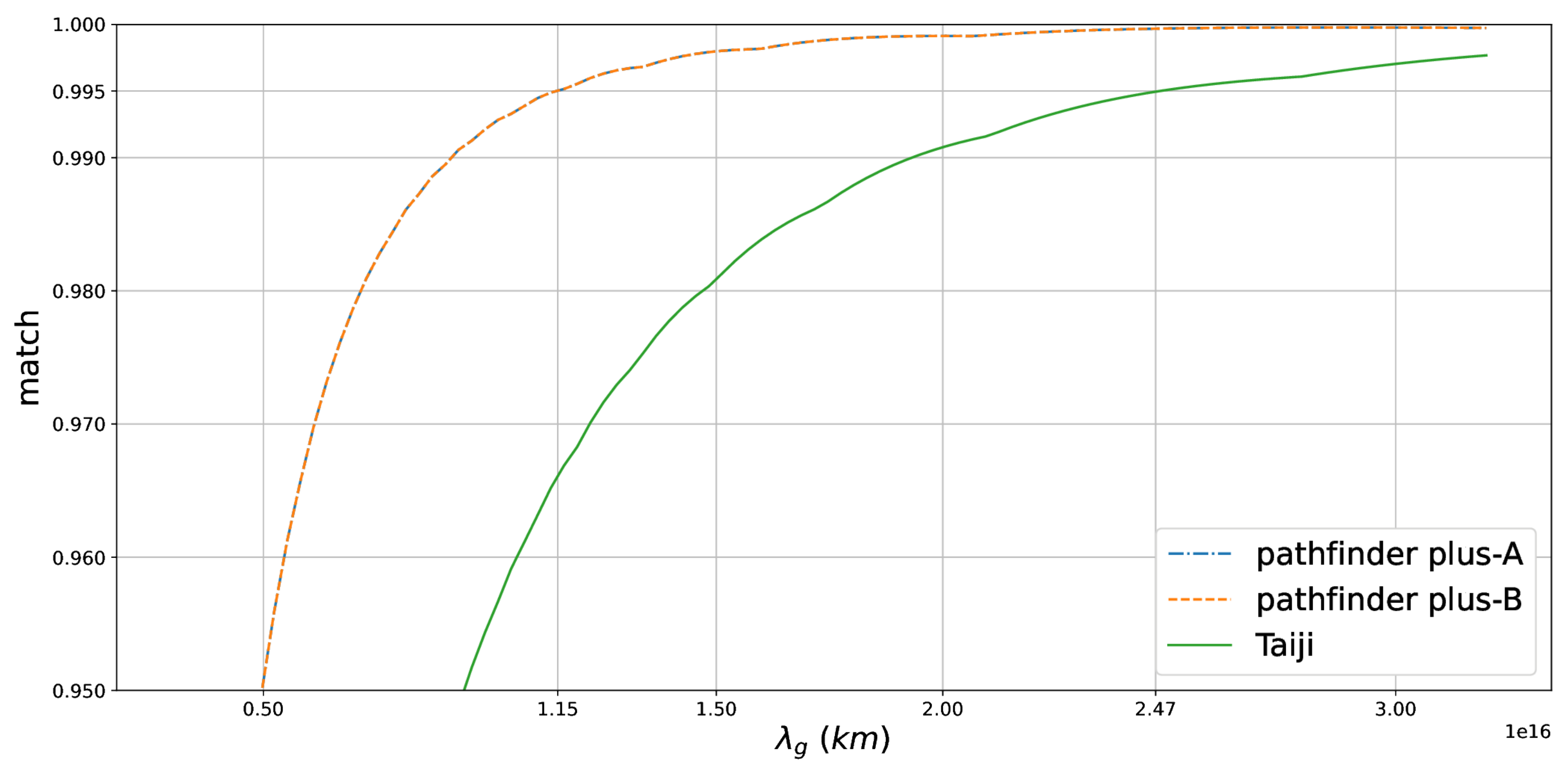}
\caption{\label{d_match} Match values between GR and dispersive waveforms in function of $\lambda_g$. The blue line, orange dotted-line and green line respectively correspond to the case of a detection by pathfinder plus-A, pathfinder plus-B and Taiji.}
\end{figure}

As in the previous section, we set the criterion value to $0.995$. Fig. \ref{d_match} shows the match between the GR waveform and the dispersion waveform for different values of the graviton Compton wavelength $\lambda_g$. In the case of a detection of the J1430+2303 source by pathfinder plus-A or pathfinder plus-B, we can constrain $\lambda_g$ to at least $1.1\times10^{16} \rm km$. For the final configuration of Taiji, expected to be much more sensitive, we found a value of $2.5\times10^{16} \rm km$. These results are again significantly better than the one obtained using the O3 data which is about $9.7 \times10^{13} \rm km$ \cite{O3} (the constrain on $m_g$ is about $1.27\times10^{-23}{\rm eV}/c^2$).

\subsection{ppE and polarization tests}
In GR, there are two GW polarizations: $h_+$, and $h_{\times}$ modes. However, alternative gravity theories can yield to the prediction of up to six GW polarizations, where the four extra modes are denoted by $h_x$, $h_y$, $h_b$, and $h_{L}$. Searching for these four additional polarizations is an important test of GR. The parameterized post-Einsteinian (ppE) framework \cite{PhysRevD.80.122003, PhysRevD.84.062003} is a model-independent framework. The basic idea behind is to introduce generic parameters in the amplitude and phase that catch non-Einsteinian effects. 

The signal observed by GW detectors is obtained by 
projecting $h_{\times}$ and $h+$ onto the detector’s antenna pattern. The response signal in the time-domain is
\begin{equation}
    h_{\rm GR}(t)=F_+h_++F_\times h_\times
\end{equation}
where $F_+$ and $F_{\times}$ are the detector antenna pattern functions for the two usual polarizations
\begin{subequations}
\begin{align}
    F_+ &= \frac{1+\cos^2\theta}{2}\cos2\phi\cos2\psi-\cos\theta\sin2\phi\sin2\psi \\
    F_{\times} &=\frac{1+\cos^2\theta}{2}\cos2\phi\sin2\psi+\cos\theta\sin2\phi\cos2\psi.
\end{align}
\end{subequations}
The ppE waveform in the frequency domain can be written as
\begin{equation}
\tilde{h}(f)=\tilde{h}_{\rm GR}(f)[1+\alpha' u^{a'}]e^{i\beta'\mu^{b'}}
\end{equation}
where $u\equiv(\pi\mathcal{M}f)$ and $(\alpha',a',\beta',b')$ are the ppE parameters used to introduce changes in the amplitude and phase. Here, the superscript $'$ is used for distinguishing the following parameters $\alpha$, $\beta$, and b, which have coefficient differences defined in \cite{PhysRevD.86.022004}.

The model-independent ppE framework with all six possible polarization modes is given by \cite{PhysRevD.86.022004}
\begin{equation}
\begin{aligned}
\tilde{h}_{\rm ppE}=&\tilde{h}_{\rm GR}(1+c\beta u_2^{b+5})e^{2i\beta u_2^b}\\&+[\alpha_bF_b\sin^2\iota+\alpha_LF_L\sin^2\iota+\alpha_xF_x\sin2\iota+\alpha_yF_y\sin\iota]\\&\times\frac{\mathcal{M}^2}{D_L}u_2^{-7/2}e^{-i\Psi^{(2)_{\rm GR}}}e^{2i\beta u_2^b}
\end{aligned}
\end{equation}
where $\mathcal{M}$ is the chirp mass, $u_2\equiv(\pi\mathcal{M}f)^{1/3}$, $\beta$, and $b$ are the free ppE parameters, and $c$ is a coefficient which depends on $b$, defined below. The parameters $(\alpha_b, \alpha_L, \alpha_x, \alpha_y)$ correspond to the breathing, longitudinal, and $x$, $y$ vector polarization modes, $(F_b,F_L,F_x,F_y)$ are the detector antenna pattern functions corresponding to each extra polarization mode. Here, c takes in account the conservative and dissipative corrections in equation of \cite{PhysRevD.86.022004}.
\begin{equation}
c=-\frac{16}{15}\frac{b(3-b)(b^2+7b+4)}{b^2+8b+9}
\end{equation}
%
\begin{subequations}
\begin{align}
    F_x &= -\sin\theta\cos\theta\cos2\phi\sin\psi+ \sin2\phi\cos\psi\\
    F_y &= -\sin\theta\cos\theta\cos2\phi\cos\psi-\sin2\phi\sin\psi\\
    F_b  &= -\frac{1}{2}\cos2\phi\sin^2\theta\\
    F_L  &=  \frac{1}{2}\cos2\phi\sin^2\theta.
\end{align}
\end{subequations}

We consider 14 parameters which are the primary and secondary SMBH masses ($m_1$ and $m_2$), the primary SMBH spin $\chi_1$, the luminosity distance $D_L$, the polar and azimuthal angles of the source's angular momentum in the ecliptic (detector) frame $\theta$, $\phi$ ($\theta_L$, $\phi_L$), and the six ppE parameters: $\beta$, b, $\alpha_b$, $\alpha_L$, $\alpha_x$, $\alpha_y$. We fix $\beta=0.01$, $b=-3$, $\alpha_b;\ \alpha_L;\ \alpha_x;\ \alpha_y=0$ in the FIM calculations, where $b=-3$ corresponds to the massive graviton theory\cite{PhysRevD.57.2061,2004CQGra..21.4367w,2005CQGra..22S.943B,PhysRevD.80.044002,2009CQGra..26o5002A,PhysRevD.82.122001,PhysRevD.81.064008} and pick the $\beta=0.01$ as the bounded result of $b=-3$ in \cite{PhysRevD.49.2658}. Table \ref{ppEt} shows the result of FIM calculations for $\Delta\lambda_i$.For Taiji pathfinder plus-B, our analysis indicates that we can measure very accurately the ppE parameters with $\Delta\beta$ at $10^{-3}$ order, $\Delta b$ at $10^{-1}$, and $\Delta\alpha_b; \Delta\alpha_L;\Delta\alpha_x; \Delta\alpha_y$ at $10^{-3}$ order. As a reference, the limits of the final Taiji will be further improved by about 2 orders.

\begin{table*}[t]
\footnotesize
\caption{Parameter estimation of ppE parameters. The results for Taiji are listed just as references.}
\label{ppEt}
\tabcolsep 21pt 
\begin{tabular*}{\textwidth}{c|c|c|c|c|c|c}
\hline\hline
 &$\Delta\beta$&$\Delta$ b&$\Delta\alpha_b$&$\Delta\alpha_L$&$\Delta\alpha_x$&$\Delta\alpha_y$\\
 \hline
pathfinder plus-B&{$7\times10^{-3}$}&{$5\times10^{-1}$}&{$5\times10^{-3}$}&{$5\times10^{-3}$}&{$2\times10^{-3}$}&{$4\times10^{-3}$}\\
\hline
Taiji&{$2\times10^{-5}$}&{$1\times10^{-3}$}&{$7\times10^{-5}$}&{$7\times10^{-5}$}&{$3\times10^{-5}$}&{$6\times10^{-5}$}\\
\hline\hline
\end{tabular*}
\end{table*}

\section{Discussion and conclusions}

Using GWs as standard sirens, pathfinder plus-B could measure the luminosity distance $D_L$ with an error just a few percent. The final Taiji configuration will improve this result by two orders. Detecting the ringdown signal of J1430+2303 merger with pathfinder plus-B, allow us to the measure the final mass and spin  with sufficient accuracy to be able to provide a very precise test on the nature of the black hole. Considering the remnant BH, if the horizon is replaced by a hard surface, we found that pathfinder plus-B can detect GW echo signal. The J1430+2303 type SMBHB merger offers a unique chance to study the nature of extremely compact objects.  In addition such a detection by pathfinder plus-B can be used to test Hawking's area theorem more accurately than the present result from the LVK collaboration. We also established that we might be able to distinguish a signal in a presence of a gravitational recoil if the kick velocity is $v_{k}\gtrsim 900$~km/s. Additionally, we were also interested to understand the benefit of such a detection to test the relation of dispersion in GR or in other modified gravity theories. We found that we can improve the constrain on the graviton Compton wavelength $\lambda_g$ by at least 2 orders of magnitude compared to the best result achieved by ground-based detectors. Finally, considering the ppE framework we determined that pathfinder plus-B is able to constrain the existence of extra polarizations. These results will be enhanced by considering the final configuration of Taiji.

The GW signal emitted by J1430+2303 is strong enough to be directly detected by Taiji pathfinder plus-A. By improving the noise level to 100 pm/$\sqrt{\rm Hz}$ and 60 fN/$\sqrt{\rm Hz}$ and building Taiji pathfinder plus-B, we will produce rich scientific results. The merger time is estimated to be about three years with uncertainties. This is just on the
edge of the Taiji pathfinder schedule.

Therefore, it is very beneficial to add an additional satellite to build Taiji pathfinder plus-A, and even more to increase the sensitivity to reach the Taiji pathfinder plus-B configuration. If an imminent merger of SMBHB can be detected at time by Taiji pathfinder plus-B, it will be the first detection of GWs produced by a binary of SMBHs. In addition, we showed that  Taiji pathfinder plus-B project is expected to yield many scientific results, such as more accurate estimation of GW source parameters, tests on the nature of BHs and gravity theory\cite{2021SCPMA..6452011Y,2022SCPMA..6500412L}. The results will be greatly improved in comparison with the currents results obtained by the LVK collaboration. Our analyses are based on the J1430+2303 source, and although there is an uncertainty in the time of the merger, all our results are relevant and generalizable to similar SMBHs when these detectors have reached their optimal performance.


\section{Acknowledgements}This work is supported by The National Key R\&D Program of China Grant No. 2021YFC2203002 and No. 2020YFC2201501, NSFC (National Natural Science Foundation of China) No. 11773059, No. 12173071, No.~12147103 and No.~11821505, and the Strategic Priority Research Program of the CAS under Grants No. XDA15021102. W. H. is supported by CAS Project for Young Scientists in Basic Research YSBR-006.

\bibliography{groupref}
\end{document}